\documentclass[prl,twocolumn,showpacs,superscriptaddress]{revtex4}%

\usepackage{graphicx}
\usepackage{dcolumn}
\usepackage{bm}

\begin{document}

\title{Line shape of the $\bm\mu$H$\bm (3p-1s)$ hyperfine transitions}

\author{D.\,S.\,Covita}
\affiliation{Dept. of Physics, Coimbra University, P-3000 Coimbra, Portugal}%
\affiliation{Paul Scherrer Institut (PSI), CH 5232-Villigen, Switzerland}
\author{D.\,F.\,Anagnostopoulos} 
\affiliation{Dept. of Materials Science and Engineering, University of Ioannina, Ioannina, GR-45110, Greece}
\author{H.\,Gorke}
\affiliation{Zentralabteilung f\"{u}r Elektronik, Forschungszentrum J\"{u}lich GmbH, D-52425 J\"{u}lich, Germany}
\author{D.\,Gotta} 
\affiliation{Institut f\"{u}r Kernphysik, Forschungszentrum J\"{u}lich GmbH, D-52425 J\"{u}lich, Germany}
\author{A.\,Gruber}
\affiliation{Stefan Meyer Institut for Subatomic Physics, Austrian Academy of Sciences, A-1090 Vienna, Austria}
\author{A.\,Hirtl}
\affiliation{Stefan Meyer Institut for Subatomic Physics, Austrian Academy of Sciences, A-1090 Vienna, Austria}
\author{T.\,Ishiwatari}
\affiliation{Stefan Meyer Institut for Subatomic Physics, Austrian Academy of Sciences, A-1090 Vienna, Austria}
\author{P.\,Indelicato}
\affiliation{Laboratoire Kastler Brossel, UPMC-Paris 6, ENS, CNRS; Case 74, 4 place Jussieu, F-75005 Paris, France}
\author{E.-O.\,Le\,Bigot}
\affiliation{Laboratoire Kastler Brossel, UPMC-Paris 6, ENS, CNRS; Case 74, 4 place Jussieu, F-75005 Paris, France}
\author{M.\,Nekipelov}
\affiliation{Institut f\"{u}r Kernphysik, Forschungszentrum J\"{u}lich GmbH, D-52425 J\"{u}lich, Germany}
\author{J.\,M.\,F.\,dos\,Santos}
\affiliation{Dept. of Physics, Coimbra University, P-3000 Coimbra, Portugal}%
\author{Ph.\,Schmid}
\affiliation{Stefan Meyer Institut for Subatomic Physics, Austrian Academy of Sciences, A-1090 Vienna, Austria}
\author{L.\,M.\,Simons}
\altaffiliation[present address: ]{Dept. of Physics, Coimbra University, P-3000 Coimbra, Portugal}%
\affiliation{Paul Scherrer Institut (PSI), CH 5232-Villigen, Switzerland}
\author{M.\,Trassinelli}
\altaffiliation[present address: ]{Inst. des NanoSciences de Paris, CNRS UMR7588 and UMPC-Paris 6, F-75015 Paris, France}%
\affiliation{Laboratoire Kastler Brossel, UPMC-Paris 6, ENS, CNRS; Case 74, 4 place Jussieu, F-75005 Paris, France}
\author{J.\,F.\,C.\,A.\,Veloso}
\affiliation{Dept. of Physics, Aveiro University, P-3810 Aveiro, Portugal}%
\author{J.\,Zmeskal}
\affiliation{Stefan Meyer Institut for Subatomic Physics, Austrian Academy of Sciences, A-1090 Vienna, Austria}

\date{\today}

\begin{abstract}
The $(3p-1s)$ X-ray transition to the muonic hydrogen ground state was measured with a high-resolution crystal 
spectrometer. A Doppler effect broadening of the X-ray line was established which could be attributed to 
different Coulomb de-excitation steps preceding the measured transition. The assumption of a statistical 
population of the hyperfine levels of the muonic hydrogen ground state was directly confirmed by the experiment 
and measured values for the hyperfine splitting can be reported. The results allow a decisive test of advanced 
cascade model calculations and establish a method to extract fundamental strong-interaction parameters 
from pionic hydrogen experiments. 
 
\end{abstract}

\maketitle

A series of experiments has been conducted at the Paul Scherrer Institut (PSI), Switzerland, to extract the isospin 
separated pion-nucleon scattering lengths from the observation of X-ray transitions feeding the ground state 
of pionic hydrogen\,\cite{PSI98,Sch01,Got08}. With X-ray energies  of about 2-3\,keV and values for the 
strong-interaction shift and broadening being of about 7\,eV  and 1\,eV, respectively, the use of a 
high resolution Bragg spectrometer was mandatory to reach the envisaged precision on the per cent level. 

The experimental difficulties are considerable, especially for the isovector scattering 
length, which is equivalent to the determination of the strong interaction broadening. 
It requires to extract from the measured line shape a Lorentzian profile representing the natural width,  
which is convoluted with the spectrometer response and several contributions 
owing to the Doppler broadenings from different high velocity states of the exotic atom.

High velocity states develop in exotic hydrogen atoms during the atomic de-excitation cascade. 
As the system is electrically neutral, it may dive deeply into the electron cloud of a neighbouring hydrogen 
molecule. Such close collisions strongly influence the atomic cascade. The most important of these processes 
are the so-called Coulomb transitions\,\cite{BF78,KPP05,PP06}, collision-induced radiationless de-excitations, 
where the released energy is shared between the exotic atom and a normal hydrogen atom as recoil partner. 
During the cascade, acceleration due to Coulomb transitions and deceleration by elastic and inelastic collisions 
compete, which results in a complex and level dependent kinetic energy distribution. 

Historically, the first evidence for high velocity states was found in the charge exchange reaction 
$\pi^{-}p\to\pi^{o}n$ with stopped pions as Doppler broadening of the time-of-flight 
(TOF) of the monoenergetic neutrons\,\cite{Czi63}. Later on detailed studies confirmed that Coulomb 
de-excitation substantially affects the kinetic energy of $\pi$H\,\cite{Bad01} and $\mu$H atoms\,\cite{Poh01} 
even at lowest densities. From the TOF spectra, several components were identified and attributed 
to specific Coulomb transitions. Hints for the influence of the Doppler broadening in X-ray transitions were 
identified in experiments measuring the strong-interaction width of the $\pi$H ground state\,\cite{Sch01,Got08}. 
A correction for the cascade-induced broadening is therefore indispensible for a proper extraction of the hadronic 
contribution to the X-ray line width. However, the high precision information from the reaction $\pi^{-}p\to\pi^{o}n$ 
cannot be directly transfered to X-ray studies. Charge exchange occurs from $ns$ states, mainly with principal 
quantum numbers $n=3-5$ of the $\pi^{-}p$ system, whereas  the initial states for K X-ray emission are $np$ levels. 
Consequently, the preceding cascade steps for the two processes are different and the Doppler contributions 
to the X-ray line shape deviate from the ones derived from neutron TOF experiments. 

As the intensities of the X-ray transitions strongly depend on the hydrogen density, there was a first approach 
to extract information about the cross sections of Coulomb de-excitation from intensity studies. Different 
processes like Stark transitions and external Auger effect, however, are overwhelming in their importance for the X-ray 
intensity compared to Coulomb de-excitation. In consequence, the kinetic energy $T_{kin}$ of the exotic hydrogen atom 
was used in earlier cascade codes (the so-called Standard Cascade Model: SCM) as a fitting parameter with values 
around $T_{kin}=1$\,eV, which explained the measured intensities with sufficient accuracy\,\cite{Bor80}. 

\begin{figure}[t]
		\includegraphics[angle=0,width=.47\textwidth]{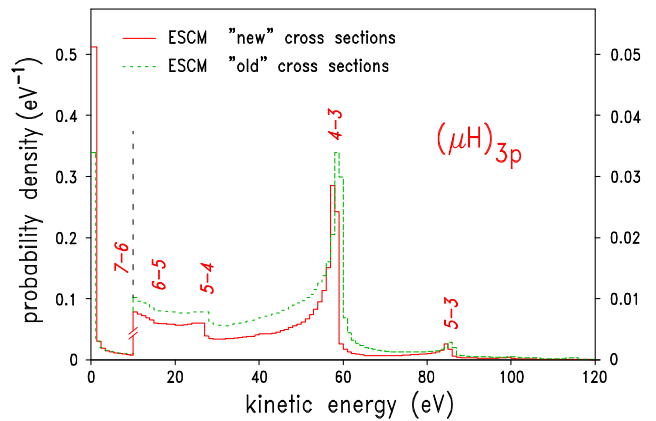}
	      \caption{
            Kinetic energy distribution of $\mu$H atoms in the $3p$ state for a hydrogen density equivalent 
            to 12.5\,bar as predicted from the ECSM\,\cite{JeMa02} using a) cross sections calculated 
            by\,\cite{JeMa02} (dashed curve) and b) recalculated cross sections\,\cite{JPP07} stimulated 
            by the present experiment (solid curve). Numbers indicate the corresponding Coulomb transitions 
            $n\,\to\,n'$. Note the change of the vertical scale at 10\,eV.}
\label{pic:Tkin} 
\end{figure}

Based on the SCM an extended standard cascade model (ESCM) was developed. It is a new approach to 
calculate the Doppler contributions to neutron TOF and exotic hydrogen X-ray spectra by taking into account 
the competing processes in each de-excitation step and a kinetic energy distribution at the time of 
X-ray emission is provided\,\cite{JeMa02}. An example of such a kinetic energy distribution for muonic 
hydrogen in the $3p$ state is shown in Fig.\,\ref{pic:Tkin}. Monoenergetic 
lines corresponding to specific Coulomb transitions $n\to\,n'$ are smeared out because 
of the numerous elastic collisions after the Coulomb transition and before X-ray emission. The validity 
of ESCM calculations cannot be tested directly in pionic hydrogen as the strong-interaction 
broadening completely masks the fine details of the various Doppler contributions. 

Muonic hydrogen as a purely electromagnetic twin system to pionic hydrogen offers itself as an ideal 
candidate for the direct observation of the Coulomb de-excitation. Ideally, it could be used as a test 
of the validity of the ESCM by reproducing the line shape of muonic hydrogen X-ray transitions in a fitting 
routine with the predictions of the ESCM as input. The theoretical values for the ground state hyperfine splitting, 
and more important, for the relative intensity of the transitions feeding the triplet and singlet components 
should be obtained. The ground-state hyperfine splitting is calculated to be $182.725\pm 0.062$\,meV\,\cite{Mar04}. 
A relative statistical population of 3:1 is expected for the triplet and singlet components. 
  
An understanding of the atomic cascade in exotic hydrogen is needed in other experimental studies as well, 
e.\,g., (i) for the precision determination of the proton charge radius from the muonic hydrogen 2$s$-2$p$ 
Lamb shift \cite{Poh01,Poh05} or (ii) in the measurement of the induced pseudoscalar coupling in muon capture 
by the proton\,\cite{Mea01,Gor04,And07}.

The experiment was performed at the $\pi$E5 channel of the proton accelerator at PSI, 
which provides a low-energy pion beam with intensities of up to a few $10^{8}$/s. Pions of 112\,MeV/c 
were injected into the cyclotron trap II\,\cite{PSI98} and decelerated using a set of degraders optimized to 
the number of muon stops by measuring X-rays from muonic helium. Muonic atoms are formed by slow muons originating from 
the decay of almost stopped pions close to or in a cylindrical cryogenic 
target cell of 22\,cm length and 5\,cm in diameter in the center of the trap. The cell was filled with hydrogen 
gas cooled down to 25K at 1\,bar absolute pressure, which corresponds to a density equivalent to 12.5\,bar at room 
temperature. X-radiation could exit the target cell axially through a 5\,$\mu$m thick mylar$^{\textregistered}$ window. 

\begin{figure}[t]
		\includegraphics[width=.47\textwidth]{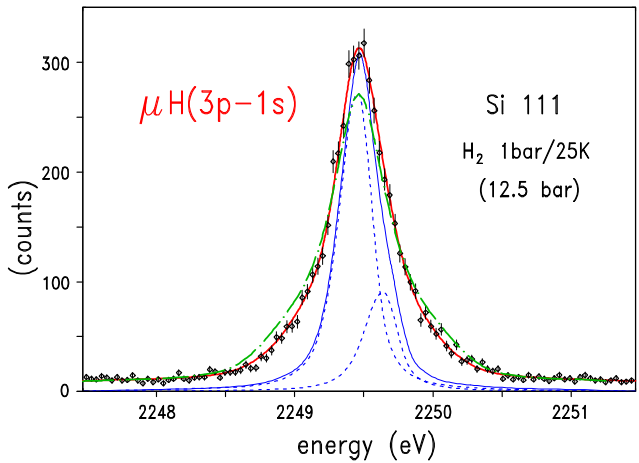}
	      \caption{Line shape of the $\mu$H$(3p-1s)$ transition as measured with a Si (111) crystal in first 
             order. One energy bin corresponds to 2 CCD pixels or 37.2\,meV. The spectrometer response (thin 
             solid line) represents the expected line shape formed by the two hyperfine components 
             (dashed lines) without any Doppler broadening (normalised to the peak height after background 
             subtraction). Including Coulomb de-excitation by using long standing 
             cross sections\,\cite{JeMa02} yields a poor description (dashed-dotted) of the line shape. The 
             ``best fit'' in a ``model free'' approach (see text) in shown 
             by thick solid line following the data. A very good fit -- being indistinguishable by eye from the 
             ``model free'' approach -- was found when using for the ESCM calculation the recently recalculated 
             cross sections\,\cite{JPP07}.}
\label{pic:muH_spectrum}
\end{figure}

X-rays from the $\mu$H$(3p-1s)$ transition were measured with a Johann-type Bragg spectrometer equipped 
with a spherically bent Si(111) crystal having a radius of curvature of $2982.2\pm 0.3$\,mm and a free diameter 
of 90\,mm\,\cite{Got04}. Such a spectrometer is able to measure simultaneously an energy interval according 
to the width of the X-ray source when using a correspondingly extended X-ray detector, in this case a 
$3\times 2$ array of charge-coupled devices (CCDs) covering in total 72\,mm in height and 48\,mm in 
width\,\cite{Nel02}. A pixel size of 40\,$\mu$m provides the two dimensional position resolution 
necessary in order to measure the diffraction image. The background rejection capability of CCDs together 
with a massive concrete shielding suppresses efficiently events from beam induced reactions. In total 
almost 10000 events were collected for the $\mu$H$(3p-1s)$ line (Fig.\,\ref{pic:muH_spectrum}). 

The spin-averaged $\mu$H$(3p-1s)$ transition energy is calculated to be $2249.461\pm 0.001$\,eV with a 
radiative line width of 0.3\,$\mu$eV\,\cite{Indpc}. The $3p$-level splittings amount to a few meV only\,\cite{Pac96}. 
Hence, two components with identical response functions are sufficient to describe the $(3p-1s)$ line. The 
spectrometer response was determined using narrow X-rays from helium-like argon as outlined in\,\cite{Ana05,Tra07}. 
Applying this method to chlorine and sulfur, an extrapolation yields a resolution of 272$\,\pm$\,3\,meV (FWHM) 
at the $\mu$H$(3p-1s)$ transition energy, which is significantly narrower than the observed line width 
(Fig.\,\ref{pic:muH_spectrum}). Details may be found elsewhere\,\cite{Covth}.

As a first trial to include the Doppler broadening caused by Coulomb de-excitation the kinetic energy distribution 
given by the ESCM result of\,\cite{JeMa02} (Fig.\,\ref{pic:Tkin}) was taken directly as an input for the fitting 
of the line shape,  which was done by means of the MINUIT package\,\cite{Jam75}. A comparison to the measured line 
shape yields a poor reduced $\chi^{2}$ of $\chi^{2}_{r}=1.353$ only (Fig.\,\ref{pic:muH_spectrum}--dashed dotted line). 

It is evident that this ESCM prediction with a weight of 36\% only for the low-energy component 
$T_{kin}\leq 2$\,eV underestimates substantially the fraction of X-rays with small Doppler shifts. 
This reveals that cross sections responsible for the development of such low energies coming from an 
interplay of Coulomb de-excitation and elastic collisions could still be deficient. In addition, the 
cascade calculation starting at $n=8$ possibly neglects stronger effects from Coulomb de-excitation in the 
outer shells ($n\,>\,11$), where the ESCM uses a classical trajectory Monte-Carlo approach. 

Consequently, as a further approach a ``model free'' method was applied, which has already been used in the case 
of the neutron TOF spectra\,\cite{Bad01}. The Doppler contributions from Coulomb de-excitation to the $\mu$H$(3p-1s)$ 
line shape were determined by modelling the kinetic energy distribution with several rectangular boxes, 
the number of which is assessed by the transitions in the preceding cascade. 

Several sets of kinetic energy boxes were investigated including besides $\Delta n=1$ also 
$\Delta n=2$ Coulomb transitions. Hyperfine splitting, i.\,e., both line positions, relative population, (flat) 
background, and the relative weight of the boxes were free parameters in the fit. The sum of the relative weights 
of the boxes was always normalized to one.

It is a major result of this study, that three narrow boxes are essential but also sufficient to model a kinetic 
energy distribution yielding a good fit to the $\mu$H$(3p-1s)$ line shape: (i) one low-energy component below 
$T_{kin}$\,=\,2\,eV collecting $\mu$H atoms which gained their energy from $n\,\geq\,10$ Coulomb transitions and/or 
high velocity systems degraded by collisions, and (ii) two at higher energies corresponding to the de-excitation 
steps  $n=5\to 4$ ($T_{kin}$\,=\,26.9\,eV) and $n=4\to 3$ ($T_{kin}$\,=\,58.2\,eV). 

A $\chi^{2}$ analysis shows that a weight of $\approx$\,60\% is mandatory for the 
low-energy component. In case of the high-energy components the fit is only sensitive to the relative weight 
and the central value. Extending the boundaries up to $\pm$\,30\% of the central values, affects the 
result by less than 1.4 standard deviations. Therefore, the kinetic energy distribution could be 
condensed\,to\,three narrow\,intervals.

The best reduced $\chi^{2}$ is found to be $\chi^{2}_{r}=0.947$ for the kinetic energy intervals set 
to [0-1.8], [26.4-27.4], and [57.7-58.7]\,eV resulting in relative weights of (61$\pm$2)\%, (25$\pm$3)\%, 
and (14$\pm$4)\% (Fig.\,\ref{pic:muH_spectrum}: ``best fit'' in the ``model free'' approach). Uncertainties 
represent statistical errors only. 

\begin{figure}[b]
		\includegraphics[width=.44\textwidth]{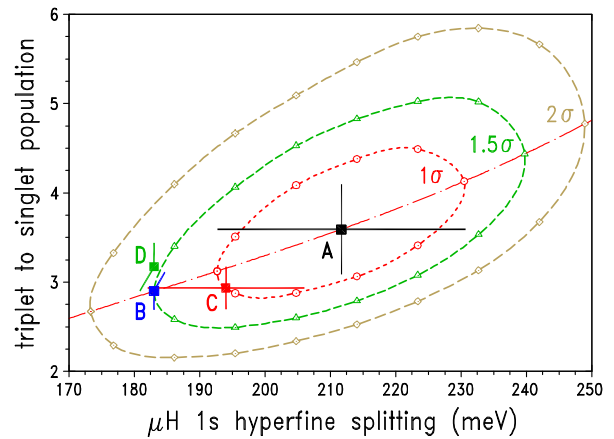}
	      \caption{$\chi ^{2}$ contour for the correlation of hyperfine splitting and relative 
            population in the ``model free'' approach. The dashed-dotted line displays the location of the 
            minimum $\chi ^{2}$ for the corresponding hyperfine splitting.
            (A -- D: see text).}
\label{pic:chi2_contour} 
\end{figure}

A correlation study of hyperfine splitting and relative population was performed by using the three kinetic 
energy intervals found in the above mentioned analysis, but with their weights, total intensity, and 
background kept as free parameters. The best $\chi^{2}_{r}=0.941$ is obtained for a hyperfine splitting of 
211\,$\pm$\,19\,meV, a triplet to singlet population of (3.59\,$\pm$\,0.51)\,:\,1 
(Fig.\,\ref{pic:chi2_contour}--A), and relative weight of $61\pm 1$\% for the low-energy component, 
where errors correspond to 1$\sigma$. The 1,\,1.5,\,and 2\,$\sigma$ contours are also shown. When fixing 
the hyperfine splitting to the theoretical value, the best fit is obtained for a triplet to singlet population 
of (2.90\,$\pm$\,0.21)\,:\,1 (Fig.\,\ref{pic:chi2_contour}--B), very close to the statistical value. The 
$\chi^{2}$ differs only by 1.5\,$\sigma $ from the best value. 

The results from the present experiment led to a reconsideration of cross sections involved in exotic-hydrogen 
de-excitation. In a fully quantum-mechanical close-coupling approach, elastic scattering, Stark transitions 
and Coulomb de-excitation now have been calculated in a unified manner\,\cite{KPP05,PP06,PP07}.

Including these cross-section results in the ESCM model\,\cite{JPP07}, the relative weight of the low 
energies ($T_{kin}\leq 2$\,eV) increases to 55\%, which is closer to the constraint found in the 
``model free'' approach. A fit to the measured $\mu$H$(3p-1s)$ line shape by using directly the new kinetic 
energy distribution (Fig.\,\ref{pic:Tkin}) yields a much better description of the data 
(Fig.\,\ref{pic:muH_spectrum}). Leaving hyperfine splitting and triplet to singlet intensity ratio as free 
parameters, values were obtained of $194\,\pm\,12$\,meV for the hyperfine splitting and (2.94\,$\pm$\,0.24)\,:\,1 
for the triplet to singlet intensity ratio (Fig.\,\ref{pic:chi2_contour}--C). The striking agreement is indicated 
by $\chi^{2}_{r}= 0.997$. When fixing the splitting to the theoretical value the relative triplet to singlet 
population becomes (3.17\,$\pm$\,0.27)\,:\,1 (Fig.\,\ref{pic:chi2_contour}--D).

Another process, which may affect the line shape, is due to molecule formation with subsequent radiative 
decay $(\mu p)_{nl} + H_{2}\to [(\mu pp)^{*}p\,ee]^{*}\to [(\mu pp)p\,ee]^{*}+\gamma$ \cite{Taq89,Jon99}. 
If radiative decay contributes significantly, a line broadening or even satellites at the low-energy side are 
expected. In the case of $\mu pp$, a branching ratio for radiative decay was predicted at the few per cent 
level\,\cite{Lin03,Kil04}. However, in this experiment, no evidence was found at the 1\% level for any 
broadening except from Coulomb de-excitation. The result is corroborated from the absense of any density 
dependence of the $\pi$H$(3p-1s)$ transition energy\,\cite{Got08}. 

To summarize, the line shape of the $\mu$H$(3p-1s)$ transition was measured with a high 
resolution Bragg spectrometer. The influence of Coulomb de-excitation was directly seen from a line broadening 
compared to the spectrometer resolution. By using a ``model free'' approach various Doppler 
contributions are identified, which are attributed to preceding Coulomb transitions. A large fraction of  
the $\mu$H systems are found to have kinetic energies below 2\,eV. The measurement 
yields the $\mu$H ground state hyperfine splitting as calculated from QED and confirms 
experimentally the statistical population of the triplet and singlet $1s$ states. The measurement triggered a new 
calculation of cross sections resulting in a significantly improved description of the $\mu$H$(3p-1s)$ line shape 
and serves as a basis for a further evaluation of the determination of the isovector scattering amplitude from 
pionic hydrogen data with increased precision.

\begin{acknowledgments}
It is a pleasure to thank V.\,P.\,Popov and V.\,N.\,Pomerantsev for the possibility to use their recent 
theoretical development. The continuous theoretical support of V.\,E.\,Markushin and T.\,S.\,Jensen is 
gratefully acknowledged.
We would like to thank B.\,Leoni, N.\,Dolfus, L.\,Stohwasser, and K.-P.\,Wieder for the technical 
assistance. The Bragg crystal was manufactured by Carl Zeiss AG, Oberkochen, Germany. Partial funding 
and travel support was granted by FCT (Lisbon) and FEDER (PhD grant SFRH/BD/18979/2004 and project 
PTDC/FIS/82006/2006) and the Germaine de Sta$\ddot{e}$l exchange program. This work is part of the PhD 
thesis of one of us (D.\,S.\,C., Univ. of Coimbra, 2008).

\end{acknowledgments}

\end{document}